\documentclass[runningheads]{llncs}

\usepackage[T1]{fontenc}
\usepackage{graphicx}
\usepackage{url}
\usepackage[font=small,skip=0pt]{caption}
\usepackage{sidecap}
\usepackage{orcidlink} 
\usepackage{wrapfig} 
\begin{document}
\title{Who Are ``We''?\\Power Centers in Threat Modeling}
%
\author{Adam Shostack\inst{}
\orcidlink{0000-0001-6837-5165}\\
 \email{shostack@uw.edu} }
\authorrunning{Shostack.}
%
\institute{University of Washington and Shostack + Associates} 

\maketitle              
\begin{abstract}
\vspace{-10pt}
I examine threat modeling techniques and questions of power dynamics in the systems in which they're used. I compare techniques that can be used by system creators to those used by those who are not involved in creating the system. That second set of analysts might be scientists doing research, consumers comparing products, or those trying to analyze a new system being deployed by a government. Their access to information, skills and choices are different. I examine the impact of those difference on threat modeling methods.

\end{abstract}
\section{Introduction}
Threat modeling is a collection of techniques for proactive security analysis of systems. The consensus industry methods are based on Shostack's Four Question Framework (``What are we working on, what can go wrong, what are we going to do about it, did we do a good job?'' \cite{Shostack-tm}) This paper builds on work by feminist scholars and activists to look at the influence of the intended users of industry methods. In other words, the use of `we' in the framework was a choice that ignored power dynamics. I suggest a threat modeling approach designed to helping people analyze a system they were not involved in creating. (Terms like `customer' or 'user' are not broad enough. Systems are often imposed, such as resume scanners, traffic cameras or border security.) For clarity, this paper avoids the convention of single author referring to themselves as `we.'

There are two main senses in which  the term \textit{threat model} is used. The earlier is `What's your threat model?' and `random oracle', or 'a network attacker,' could be  complete answers. The term was adopted into `a model of threats,' in the sense of an abstraction of possible future harms (spoofing, tampering, etc) as applied to a system under development \cite{kohnfelder}, and was deployed in informal practices such as whiteboard discussions about system security. These were adopted by  \cite{kohnfelder}, \cite{howard},  \cite{swiderski}and others into increasingly structured methodologies. The first sense is answered by a few words, the second sense is often answered with a set of diagrams, lists of threats and mititagations and tables interlinking them.

I'll refer to these approaches as `analyst' threat modeling and 'creator' threat modeling, respectively.   The first helps us understand the relevance of an attack or analysis, the second helps anticipate and thus prevent them. Interestingly, the question `what are we working on' can be applied in either, while the techniques for answering it change. Analysts  start by identifying components, data flows, and scope from a purely observational perspective. Creators have access to documentation, source code, and decision makers.\footnote{A distinction that I failed to note in a recent corporate whitepaper \cite{fourquestionwhitepaper}.}
\section{Critiques}

Sets of scholars and practitioners sought to bring creator threat modeling techniques to the analyst perspective. These included those writing under an umbrella of feminist cybersecurity and others  focused on the needs of activists. In doing so, they exposed biases and limits of the techniques. Others lacked either access to the developers, or technical knowledge of software creation or operations.

\subsection{Survey of Critiques}



Freed et al examine 'interface-bound attackers,' who cause harm while using products as intended\cite{freed2018}. Spammers, bullies, trolls, phishers and creators of deepfakes operate within system rules, yet Stamos notes these attacks caused most harm while he led security at Facebook \cite{stamos2017}.

Slupska et al attempted to threat model a smart lock, and in particular analyze it for issues of intimate partner violence (IPV) \cite{slupska2021threat}. The project exposed first, that creator perspective is limited, and second, that the techniques of creator threat modeling don't help an end user understand the problem. I'll use this as an example, because it illustrates many challenges with creator threat modeling.

Creator techniques assume a trustworthy administrator. IPV perpetrators often take control of a user session, and monitor systems for  changes. If Alice manages the lock, Bob (an abuser) may have her password or demand administrative access.  Bob may be notified if Alice limits his access. If Bob is the admin and Alice uses physical access to the lock to reset it, Bob may be notified or asked to approve the change. So how should the lock company design an access control matrix? They might focus on an admin who can create accounts or change permissions, and users who lock or unlock the door. But the use case of two users with the administrative password is unusual for computer security, and our normal response of `set an acceptable policy' may lead to a literal slap in the face. The complexity and effort of enumerating attacks may inhibit creators from investigating or recording them. If they are analyzed, the complexity of addressing them may be declared to be an `edge case' or otherwise de-prioritized. 

Additionally, creator threat modeling methods like STRIDE or kill chains don't help Alice (as an analyst) discover or reason about these problems.

Space limits our ability to discuss a growing body of work including that by Kazansky \cite{kazansky}, EFF \cite{eff},  Sterling\cite{sterling} and Levy \cite{PrivacyThreats2020}. 


\subsection{Analysis}

We can consider possible threat modelers in a space defined by technical knowledge and system knowledgem as shown in Figure 1. \begin{wrapfigure}{r}{0.5\textwidth} 
\centering
\includegraphics[width=0.45\textwidth]{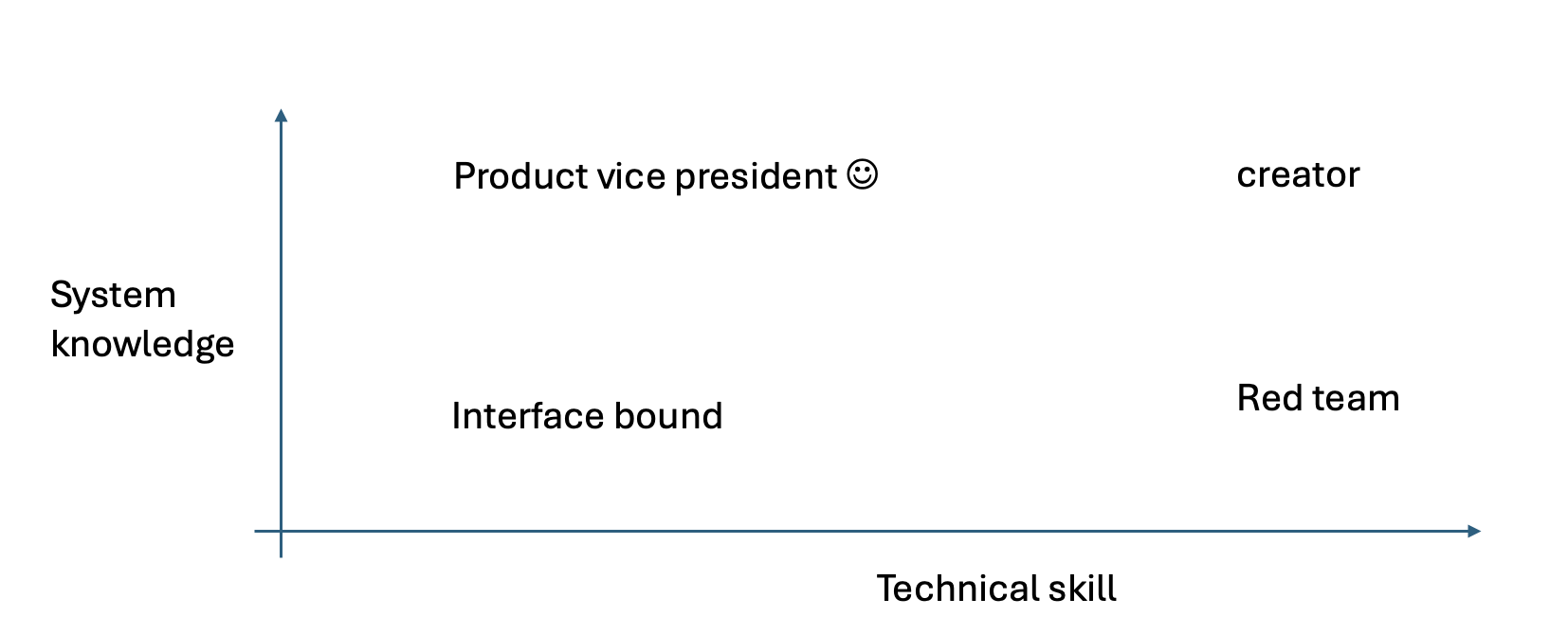} 
\caption{\small A threat modeling space}
\label{fig1}
\vspace{-30pt}
\end{wrapfigure}

\subsubsection{Social Mileu}
Microsoft recognized that design choices were being made unknowingly by developers and wanted them to be able to perform analysis. To scale, we aimed at simpler processes. (There were several downsides to this, including perhaps insufficient recognition of the quality tradeoffs between experts, and a focus on reviews and documents over skills and engagement.)  These circumstances informed the creation of threat modeling methodologies appropriate for use by technical experts to analyze systems with which they were highly familiar, or where they had access to the developers or code.\footnote{It is tempting to say easy access, but that ignores the sometimes contentious inter-team relationships.} Early versions of the Four Question Framework used 'you,' as in ``What are you working on?'', and that was intentionally changed to `we' to be more collaborative. \footnote{Other important work included that of Kohnfelder and Garg and Swiderski and Snyder. A slightly fuller history is available at \cite{fourquestionwhitepaper}.} 

This approach can be (and was!)contrasted with Anderson's educational approach.  Colleagues argued ``We can't require people to get a PhD in security,'' or ``read a 500 page book.''\footnote{Noting that the first edition of \textit{Writing Secure Code} was 501 pages including introduction, and had a quote from Bill Gates, `Required reading at Microsoft'' on the front cover.}  Anderson expected people to think critically and well,  Microsoft needed to provide a process or methodological set of steps they could follow. The focus on process was seen as a requirement for scaling, supported auditability, and was a response to a frequently expressed ``just tell me what you want me to do.''

The approach can also be contrasted to the sorts of threat modeling done by spies, attackers, bug bounty participants, or even academics who start with limited knowledge of a system, but a great deal of technical knowledge, possibly including security knowledge. They may be willing to dedicate more time, or they may see a single bug as a sufficient result. (The `single bug' goal can be contrasted with the need for creators to build a secure system.) Their technique choices and investment of energy will be shaped by those circumstances.

Microsoft's approach was an implicit decision of which participants matter. The company put technical participants (and technical threats) first. The concerns of the people impacted was not a `use case' that we discussed often. This move made perfect sense \textit{to the company}, who refered to their products as `secure by design.' This can be contrasted with the approaches required by the Food and Drug Administration, whose design-time requirements for medical device makers include a `multi-patient harm view.' Here, the FDA is acting as a counter-balancing power center relative to device makers. Deeper consideration of power relationships could improve the benefits brought by threat modeling.

\subsubsection{Technical Knowledge}
Microsoft's software engineering roles (even program management) require deep technical knowledge. Their threat modeling methodologies thus assumed technically skilled participants.

\subsubsection{Knowledge of system}
Threat modeling methodologies were developed for internal use by Microsoft product teams who were asked to engage with product security experts. Cost and effort of knowledge transfer less important because these experts would often embed for periods between weeks and years. Even so, those experts might not be briefed on features for many reasons. Those could include people doing feature work didn't see security implications, or a desire to avoid security so an insecure feature could ship. Reviews were also conducted by highly skilled experts, and likely closer to what's called product red teaming.

\section{Threat Modeling `for the rest of us'}
This section presents a simpler approach to threat modeling, designed for use by those with lower technical skill and less knowledge of a system. (The term is used for clarity, not as a judgement.)  I've selected pronouns to be  personal, even though foundational work to be done by advocates.  The Framework is:
\begin{enumerate}
\item{What have they delivered?}
\item{How will it hurt me?}
\item{Can I protect myself?}
\item{Should I even use it?}
\end{enumerate}


These questions are designed to be answerable, even if finding  answers may require specialized skills. They  aligned with the Four Question Framework to help experts remember them.  Next, I explain each question and structured approaches. These suggestions should undergo usability testing.

\subsection{What have they delivered?}
Understanding what a software package is has become more complex with the prevalence of ‘web apps’ and associated back ends, compared to earlier models of software on floppy disks.

We might be able to use a simple model of `local' and `cloud.' People believe that data on their device is private and more secure, a belief created or reinforced by both intuition, and marketing like "Your fingerprint never leaves your device." 
Questions that can be asked by those with low technical skill might include:\footnote{Usability testing these ideas is obviously important.}
\begin{itemize}
\item Does it work without internet access?
\item Can I use it without creating an account or providing a working email address?
\item What does the privacy policy tell me?
\end{itemize}

Analyzing privacy policies requires determination, and maybe skill, but can expose accessible lessons, like "We share data with our 1400 partners.'' 

Those with more technical skill browser plugins like Noscript or tools like Wireshark, and going deeper, analyst methods start to resemble those used by security researchers, rising to enumerating libraries, using a debugger or even logic probes or electron microscopy to analyze a chip or device. Firmware and mobile apps can  be downloaded and prised open, and freely available code even provides the permissions the library uses.\cite{xin2024app}

\subsection{How will it hurt me?}
Creator-oriented threat modeling may draw on frameworks like STRIDE to structure an analysis, but that requires technical skills.\cite{Shostack-threats}. A simpler set of threats, such as what does it learn and where does it send it may be helpful, but even local processing may be against the interests of a user. For example, does it show ads? Will it change function on update?

\subsection{Can I protect myself, and Should I even use it?}
The history of general-purpose computing is a history of modifying software to serve local needs, including security. Adblockers \cite{zeunert}. The trend towards restricted platforms (e.g., phones, IoT) limit user control while increasing protection against malware\cite{zittrain2008}. These restrictions complicate decisions about whether to use such systems.

More broadly, defending against trusted but untrustworthy software is challenging, even for experts. For less skilled users, it can become a Kafka-esque experience, with valid advice hard to separate from superstition.

\section{Conclusion}
The author regrets implying that threat modeling techniques are universal. Both people's depth of technical skills and their involvement in the creation of a system influence how they may threat model. 

\section*{Acknowledgements}
Julia Slupska and Leonie Tanczer helped me understand the problem they were grappling with. Josiah Dykstra, Jay Healey, Loren Kohnfelder and Kim Wuyts provided helpful feedback on drafts. Over decades, Ross Anderson's writings have profoundly influenced my own. I mourn his loss and hope to contribute this small bit to the celebration of his legacy.
 

\newpage
%
%
%
%

\end{document}